\documentclass[aps,prb,twocolumn,showpacs,amsmath,amssymb,groupedaddress,floatfix]{revtex4-1}
\usepackage{graphicx,color}
\usepackage{amsmath,amssymb,bm}
\usepackage{epstopdf}
\usepackage{braket}
\setcitestyle{square,numbers}

\newcommand{\nd}{\vphantom{\dag}}
\newcommand{\tenspr}{\hspace{-1mm} \otimes\hspace{-0.5mm}}
\newcommand{\arcosh}{\text{arcosh}}
\newcommand{\dopt}{d_{\hspace{-0.4mm}O}}

\newcommand{\nb}{n_{\text{B}}^{\vphantom{\dag}}}

\usepackage[plainpages=false,pdfpagelabels,colorlinks=true,linkcolor=red,urlcolor=blue,citecolor=blue,pdftitle={Title},pdfauthor={},pdfdisplaydoctitle=true,pdfduplex=DuplexFlipLongEdge]{hyperref}

\begin{document}
\title{Scattering of an electronic wave packet by a one-dimensional electron-phonon-coupled structure
}

\author{C. Brockt}
\email[E-mail: ]{christoph.brockt@itp.uni-hannover.de}
\author{E. Jeckelmann}
\affiliation{Leibniz Universit\"{a}t Hannover, Institut f\"{u}r Theoretische Physik, Appelstrasse 2, D-30167 
Hannover, Germany}

\begin{abstract}
We investigate the scattering of an electron by phonons in a small structure 
between two one-dimensional tight-binding leads. 
This model mimics the quantum electron transport through atomic wires or 
molecular junctions coupled to metallic leads.
The electron-phonon coupled structure is represented by the Holstein model.
We observe permanent energy transfer from the electron to the phonon system (dissipation), 
transient self-trapping of the electron in the electron-phonon coupled structure (due to polaron formation
and multiple reflections at the structure edges), 
and transmission resonances that depend strongly on the strength of the 
electron-phonon coupling and the adiabaticity ratio.
A recently developed TEBD algorithm, optimized for bosonic degrees of freedom, is
used to simulate the quantum dynamics of a wave packet launched against 
the electron-phonon coupled structure.
Exact results are calculated for a single electron-phonon site using scattering theory
and analytical approximations are obtained for limiting cases.

\end{abstract}

\date{\today}

\maketitle

\section{Introduction}
The interaction between electrons and phonons plays an important role for
the dynamical properties of solids, in particular for their electronic transport~\cite{ziman60,mahan,solyom2}.
An electron moving through a vibrating lattice can dissipate its energy, 
be scattered off course, or become dressed by a cloud of phonons,
giving rise to a quasiparticle called a polaron.
These effects are very strong in low-dimensional systems such as 
atomic wires~\cite{Agra02a,Agra02b,Ness02,Hiha12} and molecular junctions~\cite{galp07,osor08,zimb11}.
Two very recent studies~\cite{hu16,jorg16} have also demonstrated
the realization of polaronic physics in Bose-Einstein condensates of ultracold atomic gases. 
Thus various fields of physics would
benefit from a better understanding of the nonequilibrium dynamics of particles 
coupled to bosonic degrees of freedom. 

The problem of interacting electrons and phonons out of equilibrium is too complex to be solved analytically.
Thus simplified yet nontrivial many-body models are often used to obtain insights into 
the physics of these systems. 
One model of this type is the Holstein model~\cite{hols59}, that couples tight-binding 
electrons linearly to quantum harmonic oscillators, which describe the lattice vibrations. 
The equilibrium properties of this model are relatively well understood but 
studying its nonequilibrium dynamics remains a challenge, which
has attracted much attention recently~\cite{ku07,luo10,fehs11,dorf15,sayy15,mish15,zhou15,chen15}.
Indeed, strongly fluctuating bosonic degrees of freedom make analytical and even numerical studies 
cumbersome, due to the large Hilbert space dimension that must be taken into account.

\begin{figure}
\includegraphics[width=.99\columnwidth]{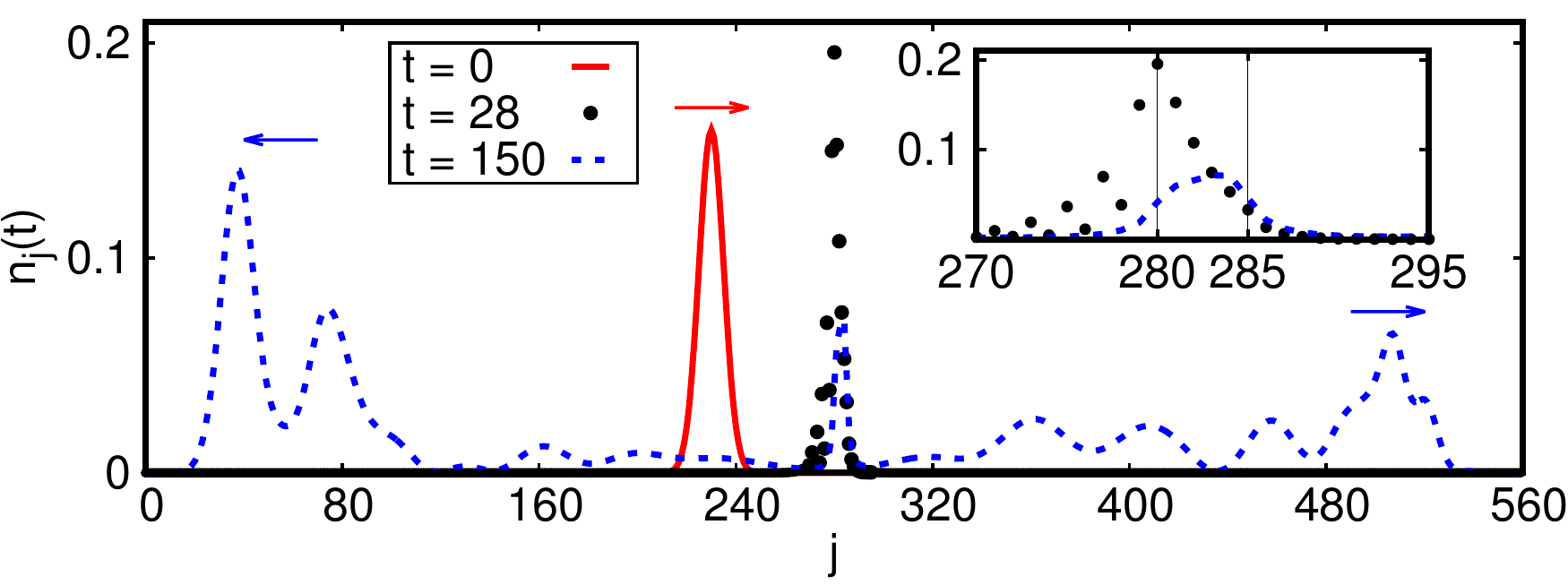}
\caption{(Color online)
The electronic density distribution calculated with TEBD-LBO for a $6$-site EPC wire
with phonon frequency $\omega_0=2.25t_0/\hbar$ and electron-phonon coupling $\gamma=t_0$ at three instances 
of time: before (red solid line), during (black dots), and after (blue dashed line) 
the main scattering processes. 
The index $1 \leq j \leq L=560$ numbers the lattice sites.
The red solid and blue dashed curves are 
multiplied by factors 2 and 10, respectively.
The inset shows an enlarged view of the region around the EPC wire.
Thin vertical lines show the position of the first and last EPC sites.
}
\label{fig:sketch}
\end{figure}

In this work we use a newly developed method~\cite{broc15} that combines the time-evolving 
block-decimation algorithm (TEBD)~\cite{vida04} with a local basis optimization 
(LBO)~\cite{zhang98} to simulate 
an electronic wave packet scattering off a phonon structure. 
In previous studies, electronic wave packets were directly injected into electron-phonon coupled (EPC) 
chains~\cite{ku07,fehs11}.
Here the EPC structure is modelled by a Holstein-type chain and is attached to long tight-binding leads at each end~\cite{broc15}. 
Its length varies between a single site (impurity) and up to 100 sites (wire).
As shown in Fig.~\ref{fig:sketch},
the electronic wave packet is initially a Gaussian distribution in the left lead with 
a momentum toward the EPC structure, where it interacts with the phonon degrees of freedom
and becomes temporarily self-trapped, and finally it is partially transmitted and reflected.

Preliminary results for this polaron scattering problem were presented in a previous work~\cite{broc15}
to demonstrate the TEBD-LBO method.
Here we extend the investigation of the rich physics offered by this simple model.
We study the time evolution of the electronic wave packet coupled to the phonon degrees of freedom.
A transient self-trapping of the electron on the EPC sites is observed due
to polaron formation and to multiple reflections at the edges of an EPC wire.
Depending on the phonon frequency and the electron-phonon coupling strength,
the wave packet can be scattered into several smaller, transmitted or reflected, wave packets
with various velocities, see Fig.~\ref{fig:sketch}.
We also study asymptotic expectation values like the 
reflection and transmission coefficients and the dissipated energy (i.e., the permanent energy 
transfer from the electron to the phonon system). We find distinct resonances in the the anti-adiabatic regime, 
which blur with decreasing phonon frequency and vanish in the adiabatic regime. 
These results can be understood using the scattering theory for one EPC impurity. 
For adiabatic weakly-coupled systems, the analysis can be extended to structures made
of several EPC sites assuming  a multiple single-site reflection ansatz.

The model and initial conditions are presented in Sec.~\ref{sec:system},
a brief summary of the TEBD-LBO algorithm is given in Sec.~\ref{sec:tebd}, 
the scattering theory for one EPC site is explained in Sec.~\ref{sec:potential_scattering},
and the method used to compare scattering theory and wave packet simulations is
detailed in Sec.~\ref{sec:packets}. 
Our numerical results are discussed in Sec.~\ref{sec:transmission} for the transmission
probability, in Sec.~\ref{sec:dissipation} for the dissipated energy, and
in Sec.~\ref{sec:transient} for the transient self-trapping.
Finally, we conclude in Sec.~\ref{sec:conc}.

\section{Problem and methods}
\label{sec:system}

We investigate a one-dimensional lattice model
that consists of three parts: a small segment of $L_H$ electron-phonon coupled sites
in the middle and two non-interacting tight-binding leads of length $L_{\text{TB}} \gg L_H$ at 
both sides. The Hamiltonian for the whole lattice reads
\begin{eqnarray}
 \label{hamiltonian}
 H & = & -t_0  \sum_{j=1}^{L-1} \left( c_{j}^{\dag} c_{j+1}^{\nd} + c_{j+1}^{\dag} c_{j}^{\nd} 
 \right) \\
 & & + \sum_{j=L_{\text{TB}}+1}^{L_{\text{TB}}+L_H}
 \left [ \hbar\omega_0 \, b_{j}^{\dag}  b_{j}^{\nd} - \gamma \left( b_{j}^{\dag} + b_{j}^{\nd} \right) 
 n_{j}^{\nd} \right ], \nonumber 
\end{eqnarray}
where
$L=L_H+2L_{\text{TB}}$ is the total number of sites while 
$b_{j}$ and $c_{j}$ annihilate a phonon (boson) and a (spinless) fermion on site $j$, 
respectively, and $n_{j}^{\nd} = c_{j}^{\dag} c_{j}^{\nd}$. 
The model parameters are the phonon frequency $\omega_0 > 0$, the electron-phonon coupling 
$\gamma$ and the hopping integral $t_0$. 

The initial state 
\begin{equation}
 \label{initialstate}
 \ket{\psi(t=0)} = \sum_{j=1}^L \psi_{j}^{\nd} c_{j}^{\dagger} \ket{\emptyset}_e \tenspr 
 \ket{\emptyset}_p, 
\end{equation}
is a tensor product of the phonon vacuum and an electronic Gaussian wave packet
\begin{equation}
\label{psivonj}
\psi_{j}^{\nd} = \sqrt{\frac{a}{\sigma\sqrt{2\pi}}} 
e^{-\frac{a^2(j-j_0)^2}{4\sigma^2} + iKja}\ ,
\end{equation}
with a width $\sigma$ sufficiently larger than the lattice spacing $a$. 
The center of the wave packet $j_0$ is always set in the left lead at $j_0=L_{\text{TB}}-10\sigma/a$.
The initial wave packet velocity $v \approx 2\frac{t_0a}{\hbar} \sin(Ka) > 0$ is set by the wave number $0<K<\pi/a$.

Our goal is to compute the time evolution of expectation values for the the state
\begin{equation}
\label{eq:dynamics}
\ket{\psi(t)} = \exp \left (-i\frac{Ht}{\hbar} \right ) \ket{\psi(t=0)} , 
\end{equation}
such as the electronic density distribution
\begin{equation}
n_j(t)  = \langle \psi(t) \vert n_j \vert \psi(t) \rangle,
\end{equation}
shown in Fig.~\ref{fig:sketch}.
The total energy 
\begin{equation}
E = \langle \psi(t) \vert H \vert \psi(t) \rangle \approx -2 t_0 \cos(Ka)
\end{equation}
is a constant of motion.
For all numerical simulations presented here, the initial variance is $\sigma^2 = 25 a^2$ and the initial wave number is $K=\frac{\pi}{2a}$
corresponding to an initial velocity $v \approx 2 a t_0/\hbar$.
Thus the average total energy is $E\approx0$. 
This corresponds to the middle of the electronic tight-binding band $-2t_0 \leq E \leq 2t_0$. 
These parameters provide a relatively broad, fast moving, and stable initial wave packet.
For all numerical data shown here, we use the energy scale $t_0=1$ and a lattice spacing $a=1$.
In addition, we set $\hbar=1$ so that the time unit is $\hbar/t_0=1$.

This system is expected to behave widely differently in different parameter regimes.
For instance, the phonon subsystem can react instantaneously (for $\hbar\omega_0 \gg t_0$) or with significant retardation 
(for $\hbar\omega_0 \ll t_0$) to the passage of the electron. In addition, it was previously determined
that dissipation can occur only for small enough phonon frequencies~\cite{broc15}.
As another example, the ratio $g=\gamma/(\hbar \omega_0)$ determines how strongly the electrons are dressed
dynamically by phonons~\cite{dorf15}.
Therefore, we are interested in all parameter regimes from the adiabatic limit $\hbar\omega_0 \ll t_0$ 
to the anti-adiabatic limit $\hbar\omega_0 \gg t_0$ and from the weak-coupling regime 
$\gamma \ll \hbar\omega_0, t_0$ to the strong-coupling regime $\gamma \gg \hbar \omega_0,t_0$.

\subsection{TEBD-LBO}
\label{sec:tebd}

\begin{figure}
\includegraphics[width=0.82\columnwidth]{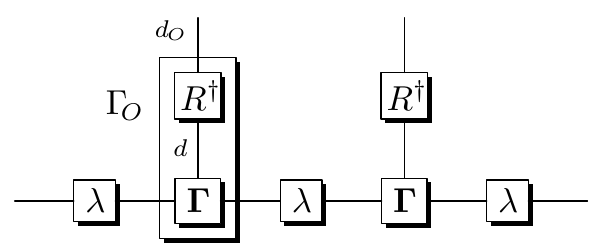}
\caption{
Schematic representation of the matrix-product state used by the TEBD-LBO algorithm.
The $\Gamma$ matrices and the diagonal $\lambda$ matrices have dimensions smaller or equal
to $D$ and have the same properties as in the original TEBD algorithm~\cite{vida04}. 
The matrix $R$ describes the local transformation from the $d$-dimensional local bare basis 
to the $d_O$-dimensional optimized subbasis. 
}
\label{fig:mps_opt}
\end{figure}

The TEBD method has proven to be a very versatile and effective numerical method
for simulating the quantum dynamics~(\ref{eq:dynamics}) in one-dimensional correlated lattice systems~\cite{vida04}.
It is based on a matrix-product-state representation of the quantum state $\ket{\psi(t)}$.
In a recent work~\cite{broc15} this algorithm was extended and optimized for large local Hilbert spaces
such as the ones needed to represent strongly fluctuating bosons. 
Instead of working in the occupation number basis (bare basis),
that is defined by $b \ket{0}_p=0$ and $b^{\dagger} b \ket{n}_p = n \ket{n}_p$,
we use the eigenbases of the single-site reduced density matrices (optimal basis) for each site~\cite{zhang98}. 
The structure of the resulting matrix-product-state is sketched in Fig.~\ref{fig:mps_opt}.
These optimized bases can change significantly in time~\cite{dorf15,broc15}, so we have to recalculate them at every time step. 
This algorithm is faster than the bare basis TEBD when the number $\dopt$ of optimal states needed to represent the quantum state
$\ket{\psi(t)}$ is much smaller than the number $d$ of required bare states. 
In that case, the computational cost drops from $\mathcal{O}(d^3 D^3)$ to $\mathcal{O}(d^3 D^2)$, 
where $D$ is the bond dimension of the matrix-product-state.

In this work we have used up to $d=386$ bare boson states in the most difficult regime, i.e. in 
the adiabatic strong-coupling regime. The number of optimized boson states reaches up to $d_O=21$
with a cutoff of $10^{-13}$
for the eigenvalues of the local reduced density matrix. We use separate 
optimized bases for a site occupied by the electron and for an empty site.
(Note that  for the case of one EPC impurity site, there is exactly one optimal mode in the
occupied-site basis because the rest of the system is always in the vacuum state.)

We always start the TEBD simulations with the lowly entangled state~(\ref{initialstate})
corresponding to $D=2$. 
The highest bond dimension reached in the simulations presented here is $D=84$
with a cutoff of $10^{-15}$
for the eigenvalues of the bipartite reduced density matrices (i.e., the squares of the diagonal 
elements of the matrices $\lambda$).
The increase of the bipartite entanglement
(and thus of the necessary  bond dimension $D$)
is mainly influenced by the length $L_H$ of the interacting structure and the
ratio $g=\gamma/(\hbar \omega_0)$. 
Thus the computational time increases faster than linear with the EPC structure size $L_H$
because of the higher bond dimensions.
(The TEBD updates for non-interacting sites take a negligible amount of time.)

We study EPC chains with up to $L_H=100$ sites in the weak-coupling regime (where
the LBO method is not needed) and
up to $L_H=6$ 
for general coupling strengths (where LBO is required), while the total system length
reaches up to $L=580$ sites.
This system length $L$ and the maximal time $t_{\text{max}}$ are tuned so that reflections of the wave packets
at the outer edges of the leads do not play any role.
As the Gaussian wave packet always starts close to the middle of the system,
the maximal time is given by $t_{\text{max}} \approx  La/(2v) \sim 10^2 \hbar/t_0$.
We typically use a time step $\Delta t = 5 \cdot 10^{-4} \hbar/t_0$ for the TEBD-LBO algorithm 
and thus $\sim 10^5$ time steps are carried out for each simulation.
The required memory 
is negligible (below 2 Gb) in all simulations but the CPU time, up to 150 hours for a single simulation, 
is a real limiting factor.

\subsection{Scattering theory}
\label{sec:potential_scattering}

For the special case of a single-site EPC impurity ($L_H=1$), we can obtain
some interesting information from the stationary scattering states.
The  stationary results should correspond to the expectation values 
of the Gaussian wave packets for asymptotic times $t \rightarrow \infty$ 
in the limit $L_{\text{TB}}-j_0, j_0  \gg \sigma/a \gg \pi/(Ka)$.

We want to solve the time-independent Schr\"odinger equation 
for the Hamiltonian (\ref{hamiltonian}) with $L_H=1$. The solution should describe an incident plane wave coming from the left
with a wave number $\pi/a > K > 0$ and being scattered by the impurity with no initial phonon excitation.
After shifting  the lattice indices so that the EPC impurity corresponds to $j=0$,  
we can write a stationary scattering state as 
\begin{equation}
 \label{scatteringstate}
 \ket{\psi}_{\text{S}} = \sum_{j,n} \psi(j,n) \ket{j}_e \tenspr \ket{n}_p,
\end{equation}
with the (bare) phonon mode index $n=0,1,...,\infty$ and the lattice site index $j=-L/2,...,L/2$. 
Far from the impurity, $\vert j\vert \gg 1$, the electronic state for a given number of phonons $n$ must be a plane wave
with a wave number $k_n$ given by the energy conservation
 \begin{equation}
\label{E_kn}
 E= -2t_0 \cos(Ka) = n\hbar\omega_0 - 2 t_0 \cos(k_n a).
\end{equation}
Therefore, the  stationary scattering state has the form
\begin{equation}
\label{eq:ansatz}
 \psi(j,n) = \begin{cases}
              A e^{i K ja}\delta_{n0} + B_n e^{-ik_nja}, & j \leq 0 \\
              C_ne^{ik_nja}, & j \geq 0
             \end{cases}
\end{equation}
with $k_0=+K$ and the amplitudes of the reflected and transmitted plane waves $B_n$ and $C_n$, respectively.
Using the uniqueness of the wave function at $j=0$, we get the conditions
\begin{equation}
 \begin{aligned}
 \label{conti}
  A + B_0 &= C_0 \\
  B_n &= C_n, \hspace{3mm} \forall\ n \geq 1.
 \end{aligned}
\end{equation}
This implies that the amplitudes for reflected and 
transmitted plane waves are equal for a given number of phonon excitations $n\geq 1$. 
(This generic result agrees with our TEBD-LBO simulations for the Gaussian wave packet.)
From~(\ref{E_kn}) we obtain 
\begin{equation}
 k_n a = \begin{cases}
              \arccos\left(\frac{n\hbar\omega_0-E}{2t_0}\right), & n < \nb, \vspace{1.2mm}  \\
              i\hspace{0.3mm} \arcosh\left(\frac{n\hbar\omega_0-E}{2t_0}\right), & n \geq \nb,
             \end{cases}
\end{equation}
where $\nb \geq 1$ is defined as the smallest index $n$ for which $n\hbar\omega_0>2t_0[1-\cos(Ka)]$. The components~(\ref{eq:ansatz}) with 
$n \ge \nb$ correspond to electronic states bound around the impurity
while the components with $\nb > n \ge 0$
correspond to scattering electronic states.
Inserting~(\ref{eq:ansatz}) with~(\ref{E_kn}) and~(\ref{conti}) in the time-independent Schr\"odinger equation, we obtain an infinite system of
recursive linear equations
\begin{eqnarray}
\label{eq:system}
  0 &= &2i\,t_0(C_0-A) \sin(k_0a) + \gamma\,C_1 \\
  0 &= & 2i\,t_0 C_n \sin(k_na) +\gamma \sqrt{n}\,C_{n-1} + \gamma \sqrt{n+1} C_{n+1}. \nonumber
\end{eqnarray}
As the normalization of the quantum state requires that $C_n \rightarrow 0$ for $n\rightarrow \infty$,
we can solve these equations backwards by setting $C_n=0$ for all $n$ larger than a high cutoff
and then verify that the results do not depend on that cutoff.
We can then calculate some stationary properties such as the transmission coefficient
\begin{equation}
 \label{T_k}
 T(K) = \sum\limits_{n=0}^{\nb-1} \frac{\sin(k_na)}{\sin(k_0a)} \left\vert\frac{C_n}{A}\right\vert^2 
\end{equation}
for an incident wave number $K=k_0$. For an elastic scattering process ($\nb=1$), one recovers
the usual result $T(K)=\vert C_0/A\vert^2$.

In the anti-adiabatic limit $\hbar\omega_0 \gg t_0$, $\nb=1$ and we can solve the equation system~(\ref{eq:system}) exactly
when the incident plane wave has an energy 
\begin{equation}
\label{eq:resonances}
E=E_m=-\varepsilon_p + m \hbar\omega_0
\end{equation}
with an integer $m \geq 0$ and the polaron energy $\varepsilon_p=\gamma^2/(\hbar\omega_0)$.
In particular, we find that
\begin{equation}
T(E=E_m) = \frac{4t_0^2-E^2}{4t_0^2}.
\end{equation}
The energies $E_m$ are equal to the eigenenergies of the EPC impurity site when it is occupied by an electron
and is disconnected from the leads. Thus the condition $E=E_m$ corresponds to a resonant tunnelling
of the incident electron with energy $E$ through the EPC impurity.

\subsection{Wave packet averaging}
\label{sec:packets}

Results such as $T(K)$ in~(\ref{T_k}) are valid for a plane wave with a sharp wave number $K$.
In our TEBD-LBO simulations, the initial electronic wave packet~(\ref{psivonj}) is a Gaussian distribution
of finite width $\propto 1/\sigma$ around $K$ in Fourier space. Therefore, we must average quantities such as transmission coefficients
over this distribution to compare the scattering theory predictions with the TEBD-LBO data.

The discrete Fourier transform of (\ref{psivonj}) is
\begin{equation}
\mathcal{F}[\psi](k) = 
\sqrt{\sigma \sqrt{\frac{2}{\pi}}} e^{-\sigma^2(k-K)^2 - ij_0(k-K)a} 
\end{equation}
where the prefactor is chosen so that $\mathcal{F}[\psi](k)$ is normalized if
$k$ is a continuous variable. This is justified by the fact that we consider the limit of infinitely long 
tight-binding leads $L \rightarrow \infty$. 
The transmission coefficient for this wave packet is then
\begin{equation}
 T^{\text{av}}(K) = \int T(k) \left\vert \mathcal{F}[\psi](k)\right\vert^2 dk.
\end{equation}
Since $T(k)$ has to be calculated through the costly solution of 
the equation system for every $k$, we approximate this integral by a discrete sum
\begin{equation}
\label{T_ps}
T^{\text{av}}(K) = \Delta k \sum_{m} \left\vert 
\mathcal{F}[\psi]\left(m\, \Delta k \right) 
\right\vert^2 T\left(m\, \Delta k \right)
\end{equation}
with intervals of length $\Delta k = \frac{\pi}{2a}\cdot 10^{-2}$.

\section{Results}
\label{sec:results}

\subsection{Transmission}
\label{sec:transmission}

\begin{figure}
\includegraphics[width=.99\columnwidth]{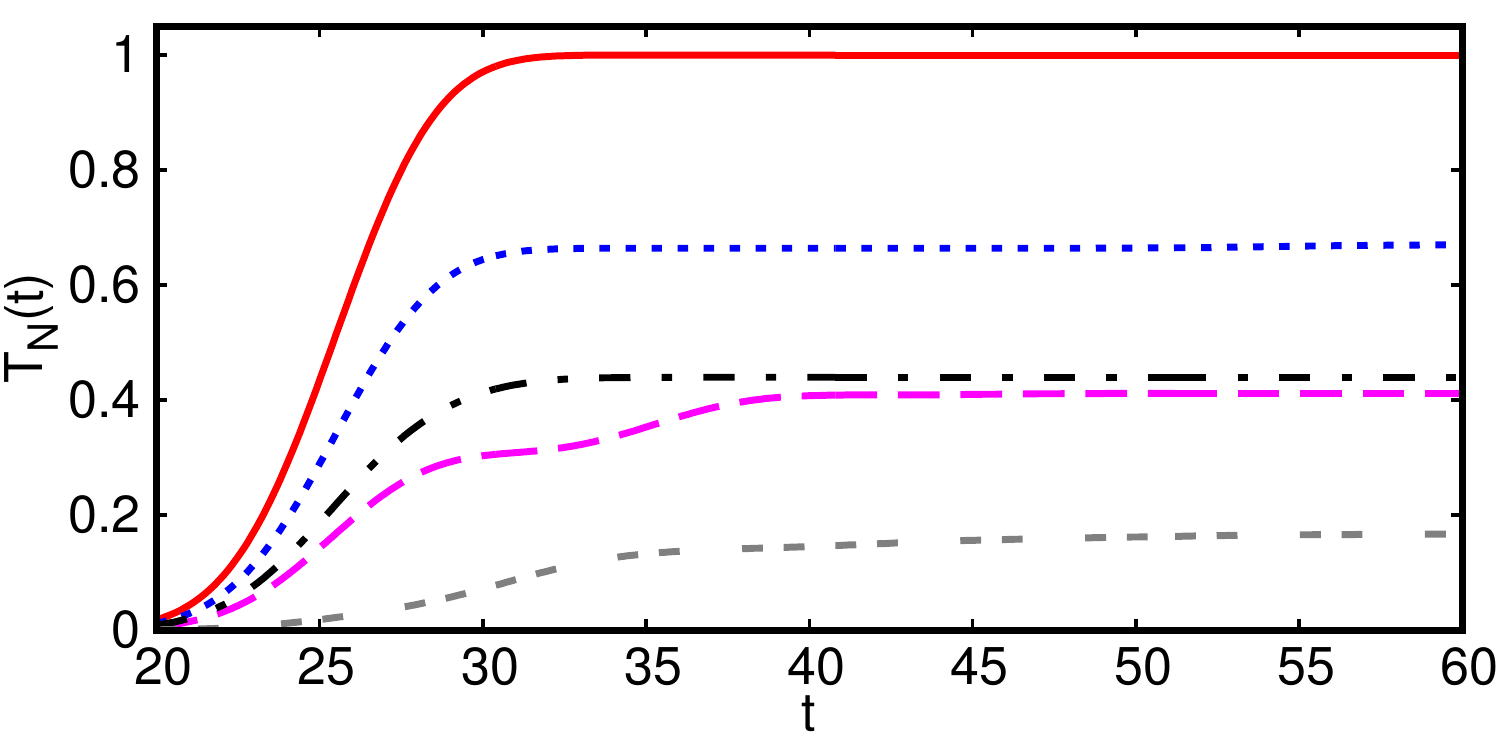}
\caption{(Color online)
Time evolution of the numerical transmission coefficient~(\ref{T_num}) calculated with TEBD-LBO for various cases:
free particle (red solid line), EPC impurity ($L_H=1$) in the 
adiabatic strong-coupling regime with $\hbar\omega_0=0.2t_0$ and $\gamma=1.9t_0$
(blue dotted line) as well as with $\hbar\omega_0=0.6t_0$ and $\gamma=3.9t_0$ (purple dashed line),
and EPC structure with one site (black dashed-dotted line) and three sites (grey double-dashed line) 
in the intermediate regime ($\hbar\omega_0=1.6t_0$ and $\gamma=1.85t_0$).
The center of the wave packet reaches the first EPC site at time $t\approx j_0a/v \approx 25\hbar/t_0$ in all cases.
}
\label{fig:transt}
\end{figure}

We first examine the transmission probability of an electron through the EPC structure.    
For TEBD-LBO simulations of Gaussian wave packets we define the transmission coefficient as the asymptotic value
of the total electronic density in the right lead
\begin{equation}
 \label{T_num}
 T_{\text{N}}(t) = \sum\limits_{j > L_{\text{TB}}+  L_{\text{H}}} n_j(t).
\end{equation}
We see in Fig.~\ref{fig:transt} that $T_{\text{N}}(t)$ seems to converge for
very long times. A similar saturation is observed for the reflected part of the Gaussian wave packet,
i.e. the total density in the left lead, 
while the probability to find the electron in the EPC structure  becomes negligibly small for long times.
Therefore, we define the transmission coefficient for each simulation of a Gaussian wave packet
as the value $T=T_{\text{N}}(t_a)$ at some large enough time $t_a$.
For a free wave packet ($\gamma=0$), we can estimate $t_a \agt (j_0a+4\sigma)/v \approx 35\hbar/t_0$ 
but for interacting systems the required times become longer as shown in Fig.~\ref{fig:transt}.
This is due to a transient self-trapping of the electron in the EPC structure, which 
is discussed in Sec.~\ref{sec:transient}.

\begin{figure}
\includegraphics[width=1\columnwidth]{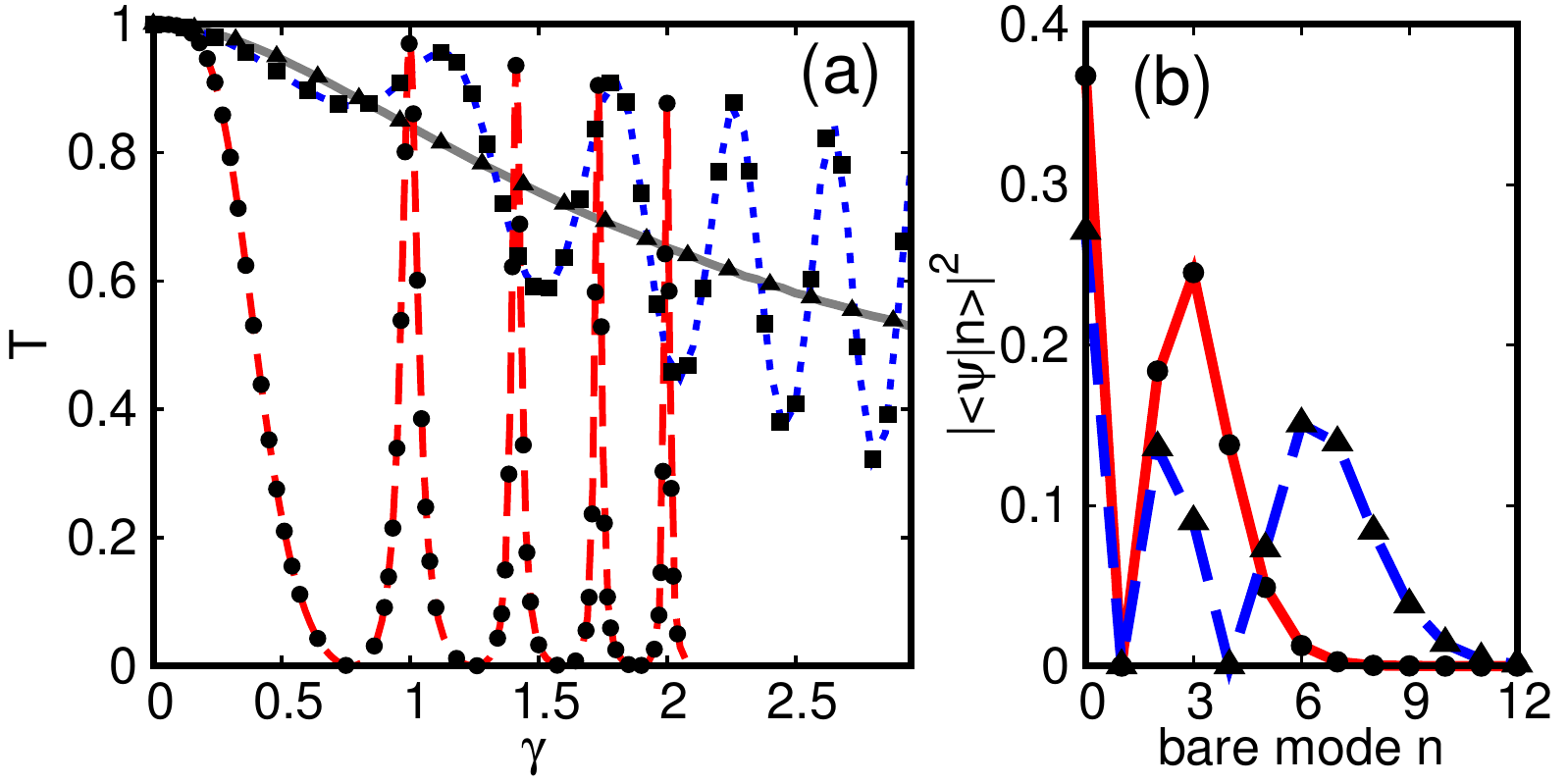}
\caption{(Color online)
(a) Transmission coefficients calculated using TEBD-LBO (lines) and with scattering theory and wave packet averaging (symbols) 
for an EPC impurity ($L_H=1$). Results are plotted as a function of the electron-phonon coupling $\gamma$  for the adiabatic regime $\hbar\omega_0=0.4 t_0$ 
(grey solid line and  triangles), the intermediate regime $\hbar\omega_0=1.35t_0$ (blue dotted line and squares), and 
the anti-adiabatic regime $\hbar\omega_0=10t_0$ 
(red dashed line and dots). 
For $\hbar\omega_0=10t_0$, $\gamma$ is divided by a factor $10$.
(b) Components $\vert \langle n \vert \psi \rangle \vert^2$ of the occupied-site optimal mode $\vert \psi\rangle$ in the bare boson basis $\vert n\rangle$
calculated with TEBD-LBO for $\hbar\omega_0 =10t_0$ and $\gamma = \hbar\omega_0$ (red solid line) 
as well as $\gamma = \sqrt{2} \hbar\omega_0$ (blue dashed line).
Also shown are the first (circles) and second (triangles) excited states of the single-site Holstein model
occupied by one electron.
}
\label{fig:trans}
\end{figure}

In Fig.~\ref{fig:trans}(a) we compare this quantity $T$ with 
the transmission coefficient (\ref{T_ps}) obtained using the scattering theory 
and the wave packet averaging for an EPC impurity.
Both approaches agree perfectly for all parameter regimes.  This confirms not only the validity
of the transmission coefficients calculated using~(\ref{T_num})
but also the overall accuracy of the TEBD-LBO simulations.

In the anti-adiabatic regime, we see clear transmission resonances and blockades as a function of
the electron-phonon coupling strength $\gamma$ in Fig.~\ref{fig:trans}(a). 
The positions of the peaks agree perfectly with the resonance condition~(\ref{eq:resonances}). Indeed,
the (average) energy of the incident electron is $E=0$ and thus Eq.~(\ref{eq:resonances}) yields the condition $\gamma = \sqrt{m} \hbar\omega_0$
with $m=0,1, \dots, \infty$.

As mentioned above, for a single-site  EPC structure the local reduced density matrix has only one 
eigenstate with finite weight when the impurity is occupied by the electron. While in general this optimal mode 
changes during time, we have found that 
it is practically constant during the whole simulation of the scattering process 
in the anti-adiabatic regime.
At the transmission peaks this quasi-stationary mode is essentially an eigenstate of
 the EPC impurity disconnected from the leads,
i.e. the single-site Holstein model occupied by one electron. The Hamiltonian corresponds to a shifted harmonic oscillator 
and its eigenstates are coherent states with the eigenenergies~(\ref{eq:resonances})~\cite{dorf15}.
Thus the optimal mode for the occupied 
impurity is the eigenstate with energy $E_m=0$. This is shown
explicitly in Fig.~\ref{fig:trans}(b), which compares the TEBD-LBO optimal modes 
and the Holstein model eigenstates
in the bare phonon basis for the peaks at $\gamma=\hbar\omega_0$ (first excited state $m=1$) 
and $\gamma=\sqrt{2} \hbar\omega_0$ (second excited state $m=2$).

This resonance mechanism can be easily understood.
As the phonon degrees of freedom are much faster than the electronic ones in the anti-adiabatic limit,
they adapt instantly to the presence of an electron in the EPC structure and thus the system
tunnels directly between eigenstates of the EPC Hamiltonian for fixed numbers of electrons.
In an extended EPC structure this implies that the phonons follow the electron ``adiabatically''
(with a small constant delay)
as shown in the Supplemental Material~\cite{suppmat} for a 100-site EPC structure 
with $\hbar\omega_0=2.9$ and $\gamma=0.25 $.

\begin{figure}[t]
\includegraphics[width=1\columnwidth]{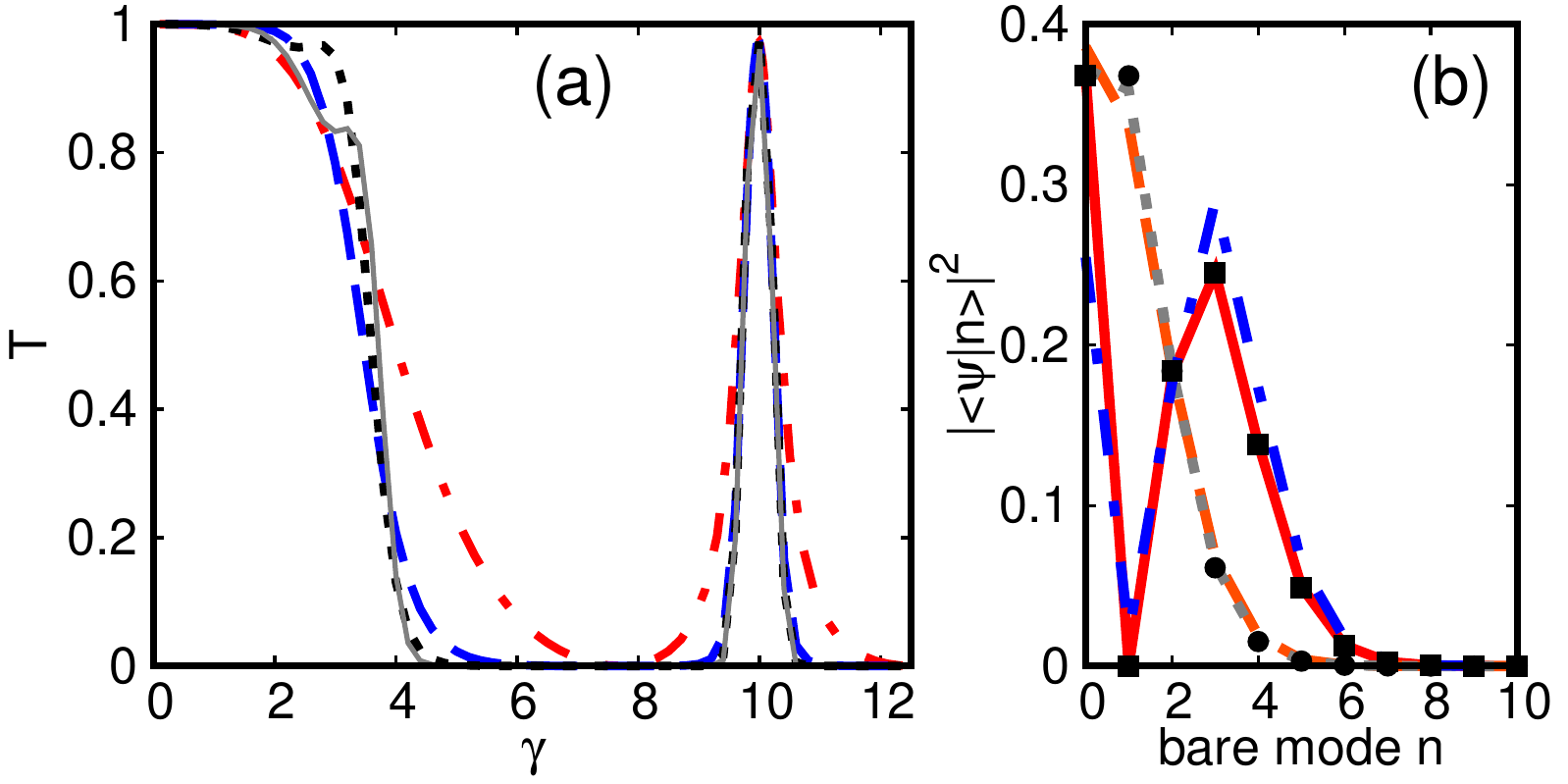}
\caption{(Color online)
(a) Transmission coefficients calculated with TEBD-LBO in the anti-adiabatic regime ($\hbar\omega_0=10t_0$)
as a function of the electron-phonon coupling $\gamma$ for several EPC wire lengths: 
$L_H=1$ (red dashed-dotted line), $L_H=2$ (blue dashed line), 
$L_H=3$ (black dotted line) and $L_H=4$ (grey solid line). 
(b) Components $\vert \langle n \vert \psi \rangle \vert^2$ 
of the two most important optimal modes $\vert \psi\rangle$ for a site occupied by an electron
in the bare boson basis $\vert n\rangle$
calculated with TEBD-LBO for $L_H=2$ and $\hbar\omega_0=\gamma=10t_0$ 
at two points in time: 
First (red solid line) and second (orange dashed line) optimal modes at $t=25\hbar/t_0$
and first (gray dotted line) and second (blue dashed-dotted line) optimal modes at $t=75\hbar/t_0$.
Also shown are the ground state (circles) and first excited state (squares) of the single-site Holstein model
occupied by one electron.
}
\label{fig:trans1234}
\end{figure}

For smaller phonon frequencies, we find smaller oscillations of the transmission coefficient as a function the 
electron-phonon coupling $\gamma$, with vanishing amplitudes in the adiabatic limit,
as shown in Fig.~\ref{fig:trans}(a).
The position of the extrema of the transmission coefficients can no longer be predicted from
the Holstein model eigenenergies.
As phonons do no longer relax instantly when the electron moves,
the optimal mode for the occupied impurity evolves in time but may still approach one of the Holstein model eigenstates 
for a finite period of time.
In addition, for smaller phonon frequencies dissipation becomes possible (see the next Section) and Eq.~(\ref{conti}) 
implies that reflection and transmission are equally probable when energy is transferred to the phonon degrees of freedom.
Thus we expect that $T\rightarrow \frac{1}{2}$ in the strong-coupling adiabatic limit.

This study can be extended to EPC wires.
Figure~\ref{fig:trans1234}(a) shows that the transmission coefficient has a similar behaviour 
for different wire lengths $L_H$ in the anti-adiabatic regime. 
Note that transmission coefficients are slightly but systematically underestimated with increasing
length $L_H$ because the probability increases that the electron is still trapped in the EPC structure
when we evaluate~(\ref{T_num}).
Interestingly, for weak coupling $\gamma < \hbar\omega_0/2$ the transmission probability does not decrease
systematically with increasing size of the EPC region. 
We see in Fig.~\ref{fig:trans1234}(a) that the largest transmission coefficient is reached 
for each wire length at some coupling $\gamma$.
We have no explanation for this surprising increase of the transmission with the EPC structure length.

The optimal modes are more involved for $L_H > 1$ than for an impurity site. 
First, the local reduced density matrix for an EPC site occupied by an electron
contains more than one eigenstate with a finite weight. Thus both the optimal
modes and their weights can evolve with time.
At the transmission resonances of the anti-adiabatic limit, however,
the most important occupied-site optimal modes seem again to approximate eigenstates of the 
single-site Holstein model. Yet their relative weights vary strongly during the scattering process.

For instance, in the strong-coupling anti-adiabatic limit the eigenenergies of the two-site Holstein model
with one electron
are approximately given by
\begin{equation}
E_{m,n} = E_m + n \hbar \omega_0
\end{equation}
where $E_m$ is the energy~(\ref{eq:resonances}) of the site occupied by the electron and
the second term is the energy of the empty site.
Thus for $\hbar\omega_0=\gamma \gg t_0$ the resonance condition $E_{m,n}=E=0$ yields either
the ground state of the occupied site ($m=0$) with an excited empty site ($n=1$)
or the first excited state of the occupied site ($m=1$) with the empty site in its ground state (the bare phonon 
vacuum state $n=0$).
Figure~\ref{fig:trans1234}(b) compares the two lowest eigenstates of the single-site Holstein model
with the two most important occupied-site optimal modes at two points in time for $\hbar\omega_0=10t_0$ and $\gamma=10t_0$.
At the start of the scattering process ($t=25\hbar/t_0$), the most important optimal state matches 
the first excited state of the Holstein model ($m=1$) while the second most important optimal state
approximates the ground state of the Holstein model ($m=0$).
After the scattering process is mostly completed ($t=75\hbar/t_0$), however, the two most important
optimal states have swapped their positions.

Therefore, in the anti-adiabatic regime we can understand  the resonance positions and identify the most important optimal states for
$L_H > 1$ like for an EPC impurity
but we cannot predict the time evolution of the weights of these optimal modes.
Outside the anti-adiabatic regime, however,
we cannot explain the (weaker) resonances and we have usually not been able to identify the optimal modes.

\begin{figure}
\includegraphics[width=.99\columnwidth]{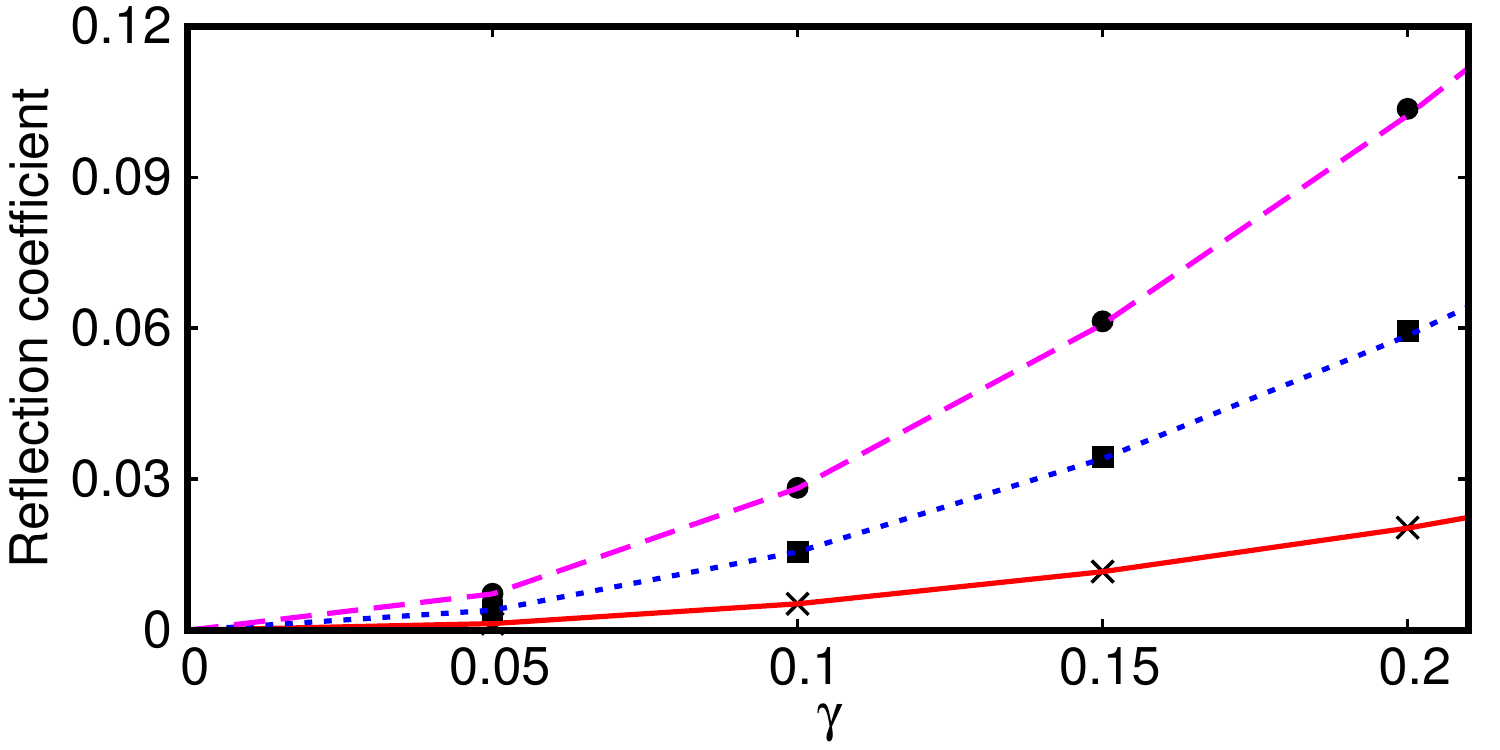}
\caption{(Color online)
Reflection coefficient $R(L_H)$ calculated with TEBD-LBO as a function of the electron-phonon coupling
$\gamma$ in the weak-coupling adiabatic regime ($\hbar\omega_0=0.6t_0$) for EPC wires
of length  $L_H=2$ (crosses), $L_H=6$ (squares), and $L_H=11$ (circles).
The lines show the predictions of scattering theory for an EPC impurity combined with Eq.~(\ref{R_n})  for 
$R(2)$ (red solid line), $R(6)$ (blue dotted line), and $R(11)$ (purple dashed line).
}
\label{fig:trans10a}
\end{figure}

Apart from the small region seen in Fig.~\ref{fig:trans1234}(a), 
the reflection is always larger for longer EPC structures. 
This is expected because the fraction of the electronic wave packet that is transmitted 
through the first site of the EPC wire can be reflected by the next one and so forth. 
In the weak-coupling adiabatic regime, the transmission coefficient $T(L_H>1)$ can be understood as 
the result of multiple independent single-site scattering processes.
We can express the reflection coefficient $R(L_H)=1-T(L_H)$ of an EPC wire
of length $L_H > 1$ 
in terms of the reflection coefficient $R(L_H=1)=1-T(L_H=1)$ of an EPC impurity
that is obtained from the scattering theory in Sec.~\ref{sec:potential_scattering} 
\begin{equation}
\label{R_n}
R(L_H) = \frac{L_H R(1)}{1+(L_H-1)R(1)}.
\end{equation}
This formula can be proven by induction.  We note that for long system lengths
$T(L_H) \propto L_H^{-1}$.
Figure~\ref{fig:trans10a} shows that Eq.~(\ref{R_n}) reproduces the results obtained with TEBD-LBO simulations
of Gaussian wave packets for EPC structures of various lengths.

It is clear that this heuristic approach cannot hold for all parameter regimes
because we assume a constant impurity transmission coefficient $T(1)=1-R(1)$, but 
its value (\ref{T_k}) is only valid if the initial phonon state is the vacuum.
After multiple scattering of the electron, however, we expect that the phonon degrees
of freedom have become excited and thus $T(1)$ will change and no longer be equal to (\ref{T_k}).
Therefore, Eq.~(\ref{R_n}) is valid only if the phonon state changes very slowly and very little, i.e., in the weak-coupling adiabatic limit.

Outside the anti-adiabatic regime or the weak-coupling regime, the computational effort
required to compute the transmission coefficients increases quickly with the EPC structure size $L_H$.
This is essentially due to the increase of bipartite and local entanglement and thus to the larger
matrix dimensions $D$ and $d_O$ required.
An additional effect is the increase of the necessary simulation time $t_a$ because the electron
stays longer partially localized in the EPC structure (see Sec.~\ref{sec:transient}).
Therefore, our results are limited to smaller electron-phonon couplings $\gamma$ or shorter lengths $L_H$.
Nevertheless, all our results suggest that the transmission for $L_H >1$ is qualitatively similar to the transmission through an EPC impurity.
For instance, the transmission coefficient for $L_H >1$ behaves
qualitatively 
as shown in Fig.~\ref{fig:trans}(a) as a function of the parameters $\omega_0$ and $\gamma$.

\subsection{Dissipation}
\label{sec:dissipation}
 
\begin{figure}
\includegraphics[width=0.98\columnwidth]{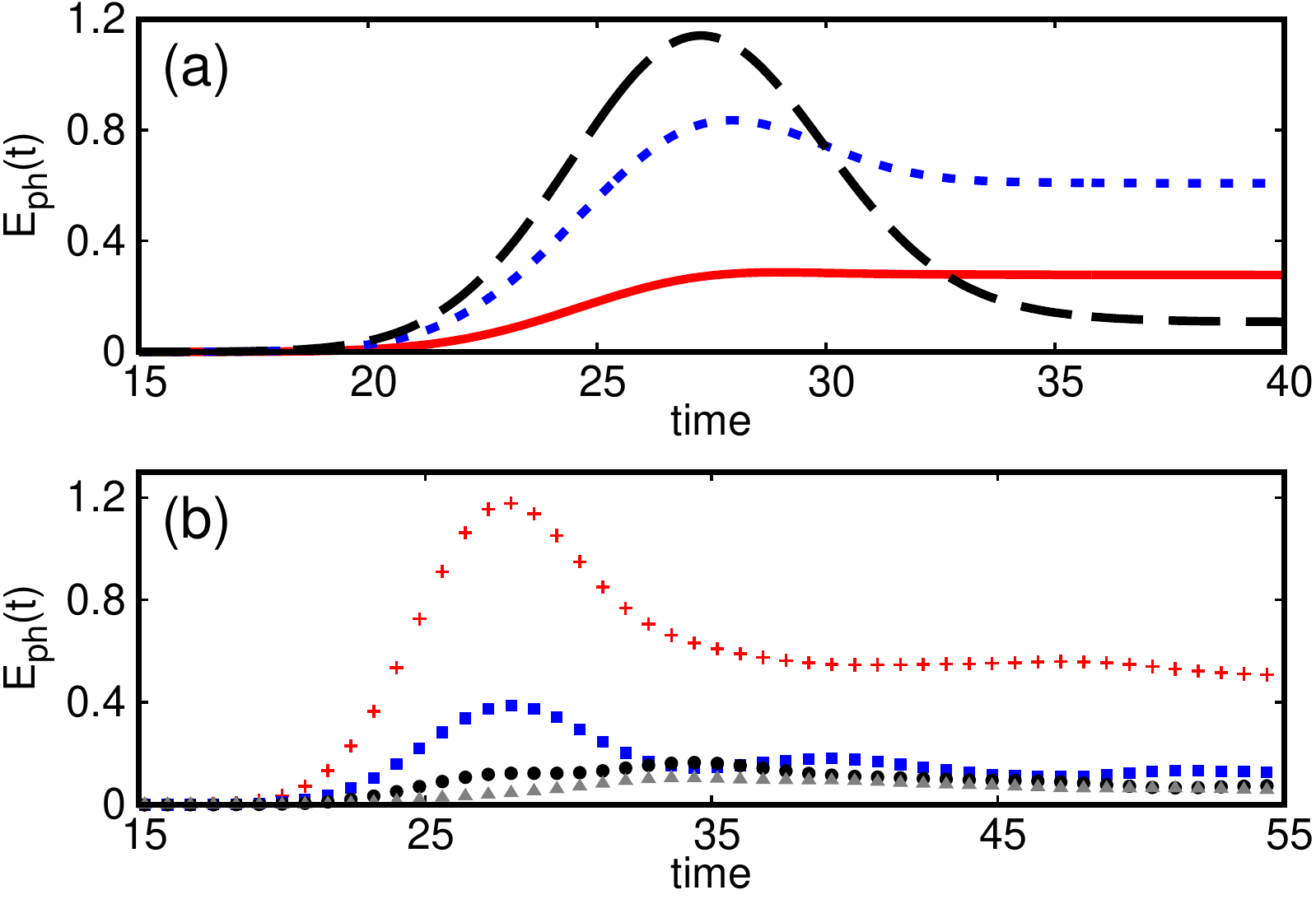}
\caption{(Color online)
Phonon energy calculated with TEBD-LBO as a function of time. (a) For an EPC impurity with $\hbar\omega_0=1.35 t_0$ and three electron-phonon couplings
$\gamma=0.8 t_0$ (red solid line), 
$\gamma=1.45 t_0$ (blue dotted line), and $\gamma=1.8 t_0$ (black dashed line). 
(b) For a four-site EPC structure in the intermediate regime ($\hbar\omega_0=1.6 t_0$, $\gamma=1.85 t_0$) the phonon 
energies are shown separately for the first (red crosses), second (blue squares), third (black bullets), and fourth (grey 
triangles) site.
}
\label{fig:e_ph}
\end{figure}

The dissipated energy $E_D$ is the energy that is transferred permanently from the
electron to the phononic degrees of freedom during the scattering process. As the electron remains only transiently
in the EPC structure (see next section), the electron-phonon interaction energy vanishes for long times. 
Consequently, $E_D$ is given by the loss of electronic energy or, equivalently, by the gain of phonon energy for long times.
Thus we evaluate the time-dependent expectation value of the phonon energy
\begin{equation}
E_{\text{ph}}(t) = \hbar\omega_0 \sum_j \langle \psi(t) \vert  b_{j}^{\dag}  b_{j}^{\nd} \vert \psi(t) \rangle .
\end{equation}
Figure~\ref{fig:e_ph} shows that this quantity converges for very long times
like the transmission coefficient~(\ref{T_num}). 
 As the initial state~(\ref{initialstate}) has no phonon energy,
we define the dissipated energy for each simulation of a Gaussian wave packet 
as the value $E_D=E_{\text{ph}}(t_a)$ at some large enough time $t_a$.
Obviously, this is the same time $t_a$ that is used to determine the transmission coefficient from~(\ref{T_num}).

Figure~\ref{fig:e_ph} also shows that the phonon energy $E_{\text{ph}}(t)$ does not increase
monotonically with time.
The behavior of $E_{\text{ph}}(t)$ depends strongly on the electron-phonon coupling $\gamma$,
as illustrated in Fig.~\ref{fig:e_ph}(a), but is also different for each site of a multi-site EPC structure, as
shown in Fig.~\ref{fig:e_ph}(b).
Nevertheless the overall behavior can be understood qualitatively.
Phonons are generated when the electronic wave packet reaches the EPC structure.
Then we observe a maximum of $E_{\text{ph}}(t)$ when the electron is mostly localized
on EPC sites. For longer times,
these phonons are annihilated when the electron leaves the EPC sites.
Yet some phonons may remain permanently excited after the electron has left the EPC structure
resulting in a permanent energy transfer from the electron to the phononic degrees of freedom.
Consequently, partial wave packets are scattered inelastically with velocities $v_n=v \sin(k_n a)$ 
that are lower than the incident velocity $v \approx 2 a t_0/\hbar$~\cite{broc15}. 

This inelastic scattering process occurs locally when the electron moves from one EPC site to the next one,
as seen in Fig.~\ref{fig:e_ph}(b). In this example, a permanent phonon 
is generated on the first EPC site reached by the electronic wave packet with a relatively high probability. 
But only transient phonons are generated on the three following EPC sites because the
initial electronic kinetic energy $2t_0$ is not high enough to generate two permanent phonons
with energy $\hbar \omega_0 = 1.6 t_0$ each.

Note that the dissipated energy cannot exceed the initial excess energy of the electron.
Thus phononic degrees of freedom with their minimal excitation energy $\hbar \omega_0$ 
cannot be permanently excited by the electron-phonon scattering process
if $\hbar\omega_0$ is larger than this initial energy~\cite{broc15}.
As the initial excess energy is always $2t_0$ in our simulations of the Gaussian wave packet,
there is no dissipation for phonon frequencies $\omega_0 \geq  2t_0/\hbar$.
In particular, there is no dissipation in the anti-adiabatic regime.

\begin{figure}
\includegraphics[width=.99\columnwidth]{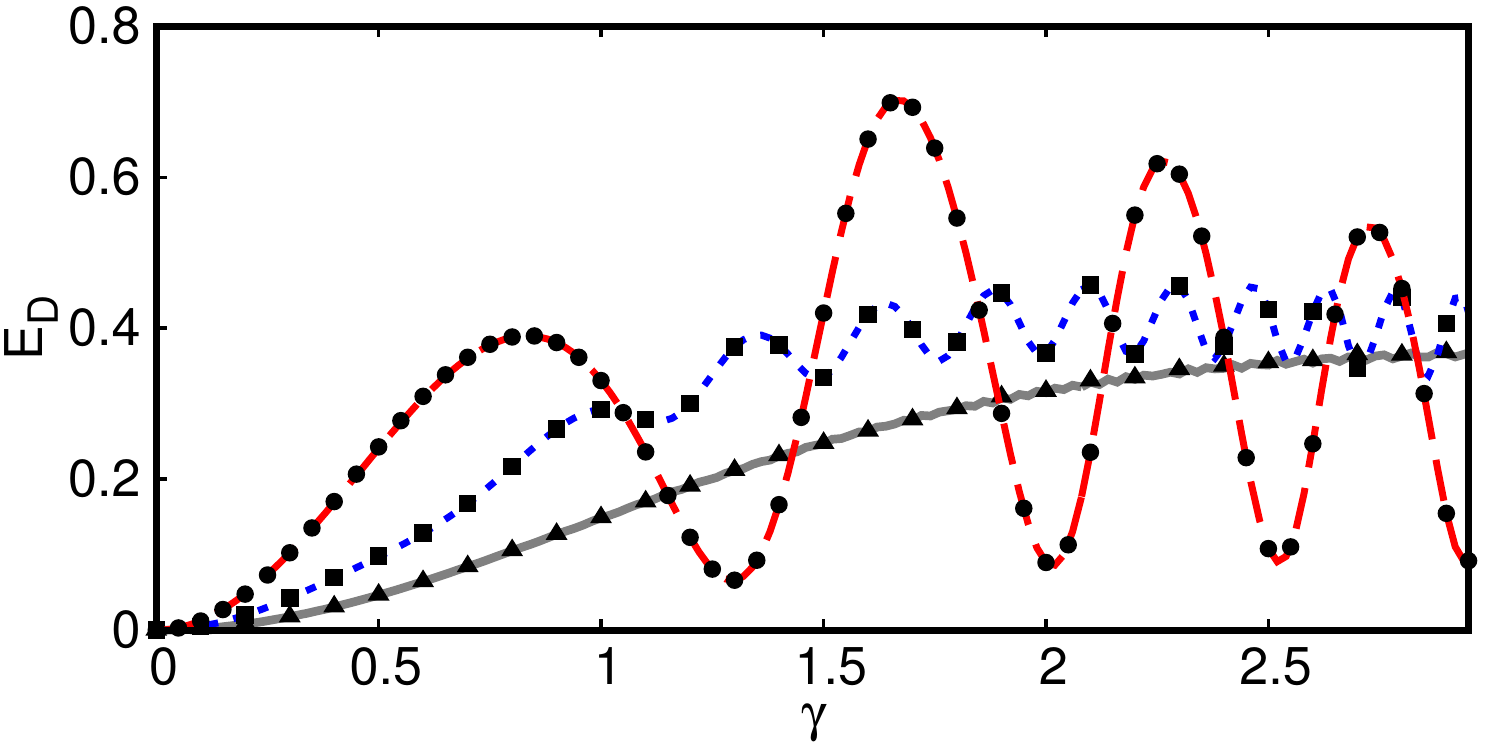}
\caption{(Color online)
Dissipated energies calculated using TEBD-LBO (symbols) and with scattering theory and wave packet averaging (lines) 
for an EPC impurity ($L_H=1$). Results are plotted as a function of the electron-phonon coupling $\gamma$  for the adiabatic regime $\hbar\omega_0=0.4 t_0$ 
(grey solid line and triangles), the intermediate regime $\hbar\omega_0=0.9t_0$ (blue dotted line and squares), and 
close to the limit of the nondissipative regime $\hbar\omega_0=1.5t_0$ 
(red dashed line and dots). 
}
\label{fig:dis}
\end{figure}

Additionally, Fig.~\ref{fig:e_ph}(a) shows that the maximal phonon energy reached during the scattering
process increases monotonically with the coupling strength $\gamma$. 
In contrast, we have found that the dissipated energy varies in a complicated way with 
$\omega_0$ and $\gamma$, as also illustrated in Fig.~\ref{fig:dis}. To shed some light on this behavior, we again examine the single-site EPC impurity
in detail.

Within the scattering theory for an EPC impurity site, we assume that the dissipated energy 
is given by the average number of excited phonons in the scattering states
\begin{equation}
 \label{E_D}
 E_{D}(K)=2\hbar\omega_0 \sum\limits_{n=1}^{\nb-1} n \frac{\sin(k_na)}{\sin(k_0a)} \left\vert\frac{C_n}{A}\right\vert^2.
\end{equation}
The factor 2 originates from the condition $B_n=C_n$ for $n>0$. 
As $\nb=1$ for $K=\pi/(2a)$ and $\hbar\omega_0 >  2t_0$, we obtain $E_{D}(K=\pi/(2a))=0$ and thus the scattering theory
confirms that there is no dissipation in the parameter regime $\hbar\omega_0 >  2t_0$.
To compare
with the TEBD-LBO simulations, we have again to average 
the quantity~(\ref{E_D})  over the Gaussian wave packet as explained for the transmission coefficient in Sec.~\ref{sec:packets}.

Figure~\ref{fig:dis} compares $E_{D}^{\text{av}}(K)$ and the energy dissipation $E_D$ calculated with TEBD-LBO
for Gaussian wave packets. The perfect agreement between both approaches confirms again the
accuracy of our TEBD-LBO method but also demonstrates that the definition~(\ref{E_D}) is correct for the dissipated energy 
within the scattering theory.
For weak electron-phonon coupling we find that the dissipation increases monotonically with $\gamma$.
 We see in Fig.~\ref{fig:dis} however that the dissipated energy oscillates for larger values of 
$\gamma$, with vanishing amplitude in the adiabatic limit.
This is reminiscent of the transmission oscillations in Fig.~\ref{fig:trans}.
Neither the extrema positions nor the oscillation frequencies (as a function of $\gamma$)
seem to coincide, however.

Nevertheless, in the special case $t_0 < \hbar \omega_0 < 2t_0$ ($\Rightarrow n_B=2$), the definition~(\ref{E_D}) and the 
scattering theory in Sec.~\ref{sec:potential_scattering} yield a simple relation between transmission coefficient $T(K)$ 
and dissipated energy $E_D(K)$ for the EPC impurity
\begin{equation}
\label{eq:reflection_dissipation}
E_D(K) = 2 \hbar \omega_0 [1-T(K)] \frac{\sin(Ka)\sin(k_1a)}{\frac{\gamma^2}{4t^2_0}+\sin(Ka)\sin(k_1a)} .
\end{equation}
This equation demonstrates that dissipation is proportional to reflection, at least in this special case, 
although the dependence of both quantities
on the model parameters $\gamma, \omega_0$, and $K$ is different.

As for the transmission coefficient, the TEBD-LBO study of the dissipated energy can be extended to multi-site EPC structures.
Again the computational effort
required to compute $E_D$ increases quickly with the EPC structure size $L_H$ outside the weak-coupling regime.
Therefore, our results are limited to smaller electron-phonon couplings $\gamma$ than for
the EPC impurity or to short lengths $L_H$.
Nevertheless, all our results suggest that the dissipation for $L_H >1$ is qualitatively similar to the dissipation through an EPC impurity.
For instance, $E_D$ behaves qualitatively 
as shown in Fig.~\ref{fig:dis} as a function of the parameters $\omega_0$ and $\gamma$.
Moreover, transmission $T$ and dissipated energy $E_D$ seems to be related in some complicated way, as discussed above.

\subsection{Transient self-trapping \label{sec:transient}}

\begin{figure}
\includegraphics[width=0.99\columnwidth]{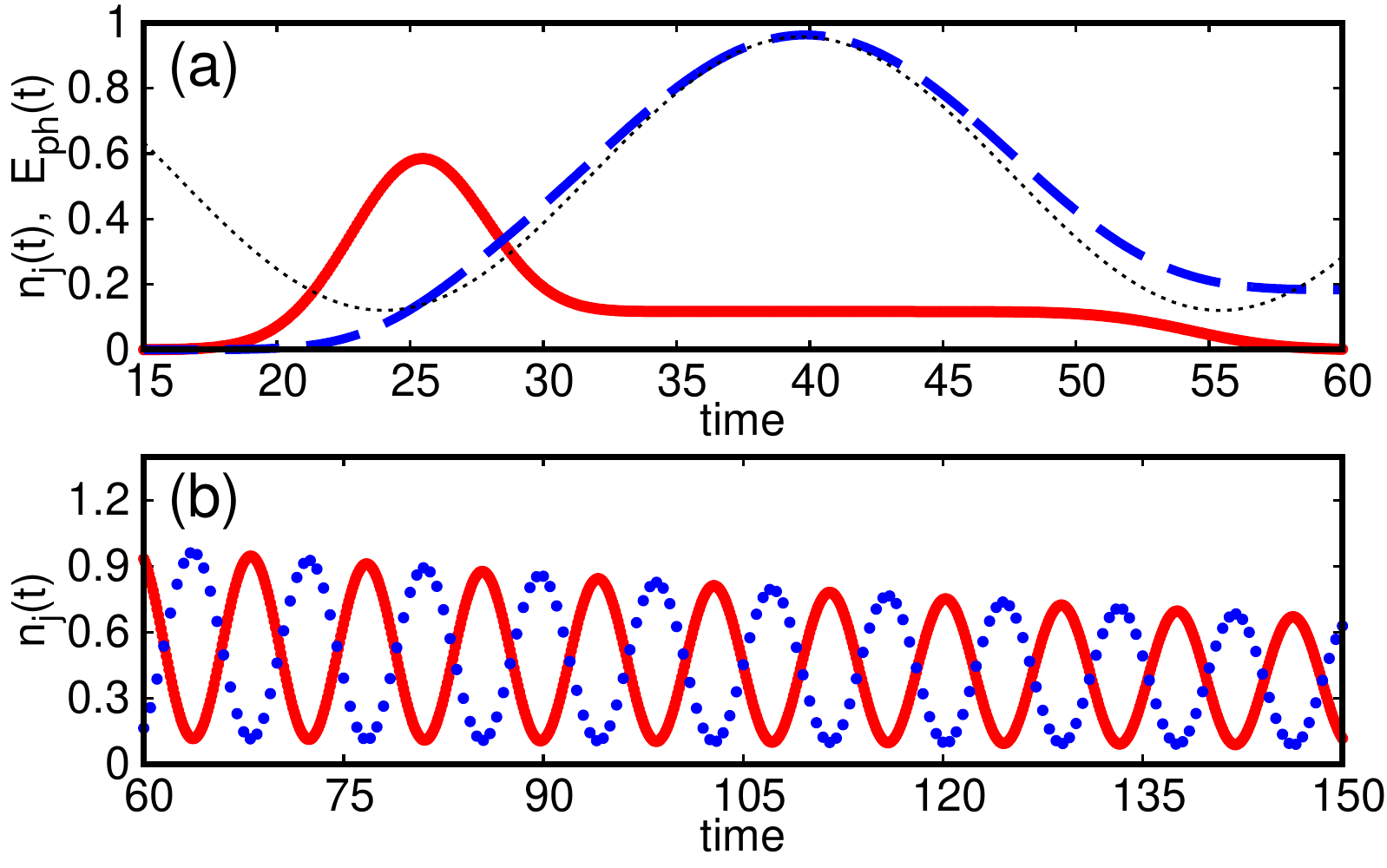}
\caption{(Color online)
(a) Electronic density on the EPC impurity (red solid line) and phonon energy (blue dashed line) calculated with TEBD-LBO
as a function of time for $L_H=1, \hbar\omega_0=0.20 t_0$, and $\gamma=1.90 t_0$.
The electron density is multiplied by a factor 10. 
The black dotted line
shows Eq.~(\ref{eq:coherent}) with $t_{\text{off}}=24 t_0/\hbar$, $E_{\text{off}}=0.12t_0$, and a 
constant average density $n_j(t=40\hbar/t_0)=0.0116$.
(b) Electronic densities (multiplied by a factor 100) calculated with TEBD-LBO 
as a function of time 
on the left (red line) and right (blue dots) sites of a two-site EPC structure 
with $\hbar\omega_0=\gamma = 10 t_0$. 
}
\label{fig:trap}
\end{figure}

In a preliminary study~\cite{broc15}, we reported that a fraction of the electronic wave packet could become temporarily
self-trapped in the EPC structure and then be belatedly transmitted or reflected.
We have verified that this transient self-trapping is quite common in the scattering problem discussed here, 
although it does not occur for all values of the model parameters.
For instance, it is responsible for the occurrence of plateaus in the time-dependent transmission coefficients
for some (but not all) model parameters, as seen in Fig.~\ref{fig:transt}.
Below we examine some aspects of this phenomenon more closely.

Figure~\ref{fig:trap}(a) shows a clear example of the transient self-trapping of the electron. 
The electronic density measured on the EPC impurity reveals that
the wave packet reaches it at time $t \approx 20 \hbar/t_0$, then  
most of the electronic wave packet has left the EPC site by $t \approx 30 \hbar/t_0$
but a small fraction remains there up to $t \approx 50 \hbar/t_0$.
The first time scale corresponds simply to the passage of the Gaussian wave packet
over the EPC site with negligible scattering and is set by the wave packet width $\sigma$ and velocity $v$ (see Sec.~\ref{sec:system}).

The second time scale is due to the trapping of the electron by the effective on-site potential 
that its presence has induced. 
The self-trapping time scale ($\Delta t \approx 30\hbar/t_0$) seems to be related to the period of the phonon degrees of freedom, 
$\frac{2\pi}{\omega_0} = 10\pi \hbar/t_0$.
During that time interval the phonon system behaves like in the single-site Holstein model
occupied by one electron, i.e, like a coherent state in a shifted harmonic oscillator. A similar behavior
was found for the decay of a highly excited charge carrier in the one-dimensional Holstein model~\cite{dorf15}.
Thus one can estimate that
\begin{equation}
\label{eq:coherent}
E_{\text{ph}}(t) \approx n_j(t) 2 \varepsilon_p^{\phantom{\dagger}} \{1-\cos[\omega_0 (t-t_{\text{off}})]\} + E_{\text{off}}
\end{equation}
where $n_j(t)$ is the fraction of the electron density that is currently trapped 
on the EPC site, $t_{\text{off}}$ is a time offset that depends on the
initial conditions (e.g., $t_{\text{off}} \approx j_0 a/v = 25 t_0/\hbar$), and 
$E_{\text{off}}$ is an energy offset that is related to the
asymptotic phonon energy $E_{\text{ph}}(t_a)$. 
Figure~\ref{fig:trap}(a) confirms that the phonon energy agrees with this equation
(using fitted parameters $t_{\text{off}} = 24 t_0/\hbar$ and $E_{\text{off}}=0.12t_0$) during
the transient self-trapping time.
Therefore, in that time interval the system has formed a highly excited polaron (a mobile quasi-particle made of an 
electron dressed by a phonon cloud) as found for the decay problem in Ref.~\cite{dorf15}.
Note that the fraction of the wave packet that gets trapped, $n_j(t)$, depends mostly on the electron-phonon
coupling strength $\gamma$.

In Fig.~\ref{fig:trap}(a) there is a single plateau in the electronic density indicating 
the generation of just one delayed partial wave packet each for transmission and reflection.
For longer EPC structures we found previously that several partial wave packets can be
transmitted and reflected at approximately equidistant times [see Fig.~3(a) in Ref.~\cite{broc15}].
The transient behavior of EPC structures with $L_H > 1$ is indeed more intricate than for
a single-site EPC impurity because there is a second self-trapping mechanism besides the polaron formation.
The electron induces an extended potential well in the EPC structure. Its wave function
can be (partially) reflected multiple times at the edge of this potential well. Thus part of the
wave packet oscillates 
inside the EPC structure and slowly leaks out into the leads.

This effect is illustrated  in Fig.~\ref{fig:trap}(b) for a two-site EPC structure 
at times $t \geq 60 \hbar/t_0$ when 99\% of the wave packet has already been transmitted or reflected.
Nevertheless, we see that a small fraction of the electronic density is still trapped inside
the EPC structure and jumps back and forth between both sites with a slowly decaying amplitude.
At every period $t\approx 8 \hbar/t_0$ a small wave packet is belatedly transmitted or reflected.
(This indicates that the wave packet velocity is about $\frac{t_0 a}{4\hbar}$ inside the EPC structure
and thus about 8 times slower than for a free wave packet.)
Due to the multiple reflections, however, the overall self-trapping time is extended greatly and 
it is clearly over
$\Delta t=150 \hbar/t_0$ in the example of Fig.~\ref{fig:trap}(b).
Finally, note that this multiple scattering at the EPC structure edges should not be confused with the
multiple inter-site scattering leading to Eq.~(\ref{R_n}).

\begin{figure}
\includegraphics[width=.99\columnwidth]{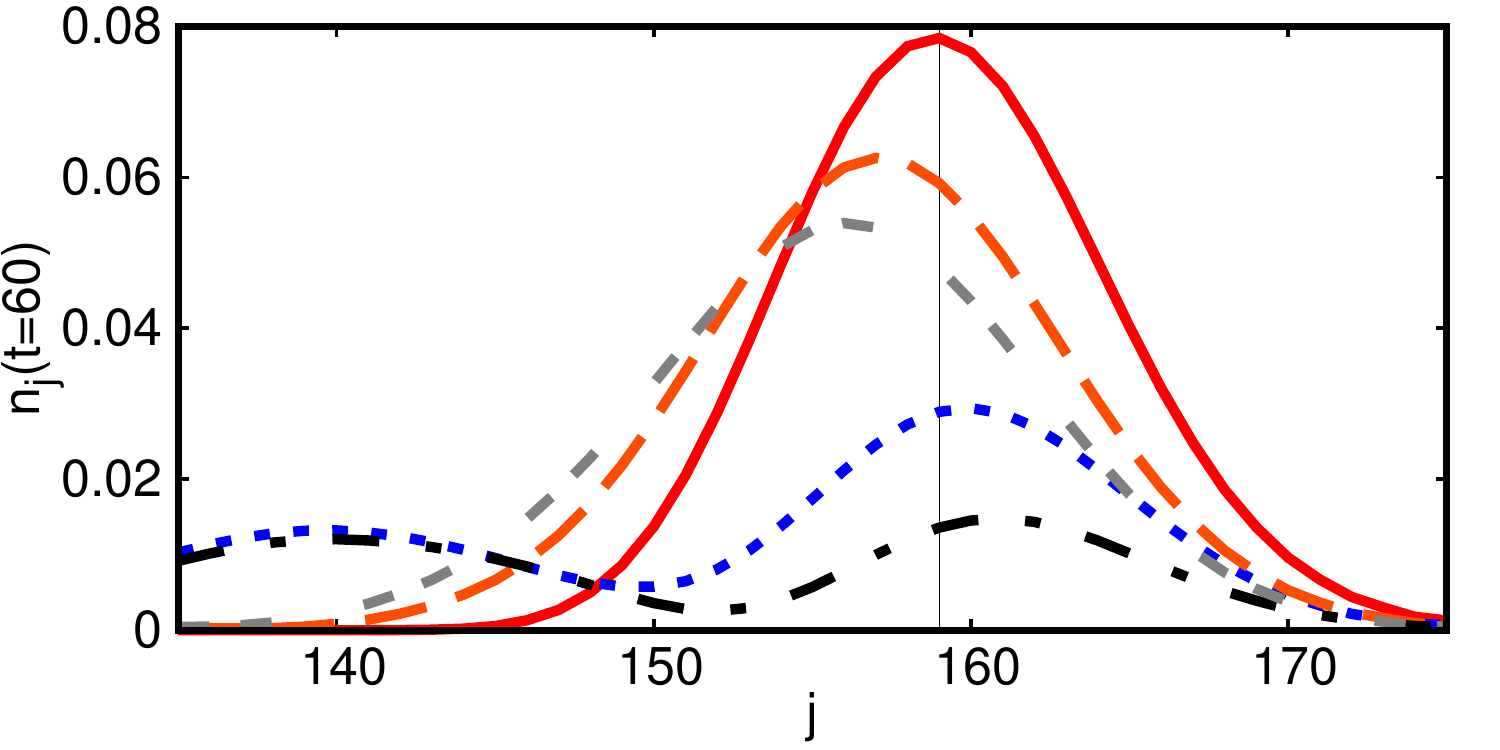}
\caption{(Color online)
The rightmost part of the electronic density distributions $n_j(t)$ calculated with TEBD-LBO at $t=60 \hbar/t_0$ 
as a function of the lattice site index $j$ for $L_H= 1$, $\hbar\omega_0=1.35 t_0$ and
several electron-phonon couplings corresponding to 
transmission minima at $\gamma= 1.5 t_0$ (blue dotted line) and at $\gamma= 2.45 t_0$ (black dashed-dotted line)
as well as to transmission maxima at $\gamma=1.80 t_0$ (orange dashed line) and at $\gamma=2.65 t_0$ 
(grey double-dashed line).  
The vertical black line shows the position of the maximum of the free Gaussian wave packet (red line).  
}
\label{fig:v135}
\end{figure}

It is tempting to define an overall self-trapping time as the delay of the wave packet transmitted
through the EPC structure compared to a non-interacting wave packet. 
For instance, one could attempt to use the distance to the rightmost maximum in the density distribution. 
We see in Fig.~\ref{fig:v135}, however, that this definition can result in a negative self-trapping time.  
For transmission minima in particular, the scattered wave packet seems to be ahead of the free one.
This is related to the Hartman effect~\cite{hartman62}: the transmission time 
of a (Gaussian) wave packet through a potential barrier can be shorter than the time
required by a free wave packet to travel a distance equal to the barrier width. 
We observe a similar effect in Fig.~\ref{fig:v135} although the scattering is not caused by a static 
potential barrier but by dynamical degrees of freedom.

This effect reveals that one has to be careful when measuring times for quantum systems. 
For a wave packet tunneling through a potential barrier, it was shown that there is a trade-off between 
the speedup of the transmitted wave packet and the damping of its amplitude~\cite{ruschhaupt09}.
As a result, the probability for the electron to reach a certain distance before some given time
is always reduced by the tunneling through a potential barrier compared to the probability 
for a free wave packet.
Our results, as illustrated in Fig.~\ref{fig:v135}, agree with Ref.~\cite{ruschhaupt09}.
Clearly, the probability of finding the electron on the right of the vertical line
is larger for the free wave packet than for any of the scattered ones although 
the density maxima are further on the right in some cases.

In all our TEBD-LBO simulations we have never observed a permanent trapping of the electron
in the EPC structure. On the one hand, this absence of localization is expected because
a trapped electron cannot dissipate its energy in the present model. Any emitted phonon will be sooner or later
reabsorbed if the electron remains long enough in the EPC structure and 
electronic energy can only be dissipated if the electron moves away in the noninteracting leads.
On the other hand, one intuitively expects that the electron should progressively lose its
kinetic energy to the phonons left behind and ultimately come to rest if the EPC wire is long enough,
like a particle in a viscous environment. Actually, once the electronic wave packet is well inside a long EPC chain, the tight-binding leads 
are no longer relevant and the relaxation dynamics should be similar to the problem of an excited electron in
the one-dimensional Holstein model, which was thoroughly investigated in Ref.~\cite{dorf15}.
Thus one expects that the kinetic energy of the electron (or the resulting polaron) should 
be permanently transferred to the phonon system in the adiabatic regime.
Unfortunately, we have not yet been able to simulate the conditions where this could happen, i.e., 
long enough EPC chains, low phonon frequency, and high dissipation rate (which implies a strong enough electron-phonon coupling)
because of the very high computational effort required.

\section{Conclusion}
\label{sec:conc}

We have studied a Gaussian electronic wave packet scattering off a one-dimensional electron-phonon-coupled  
structure connected to tight-binding leads using the TEBD-LBO method. 
We have found that this simple nonequilibrium problem offers a rich physics including
transmission resonances, dissipation, and transient self-trapping due to polaron formation and 
multiple reflections at the EPC structure edges.
Many of theses features can be understood qualitatively using scattering theory for a single-site EPC impurity.
For asymptotic expectation values (transmission coefficients and dissipated energy), we find a perfect agreement between
our numerical TEBD-LBO simulations and the exact scattering theory results.
This confirms the reliability of the TEBD-LBO method even for long-time simulations.
Nevertheless, further investigations are required to understand long EPC chains more thoroughly.
Although we discuss only the case of an incident wave number $K=\pi/(2a)$ in this paper, the
results remain qualitatively similar for other wave numbers $0 < K < \pi/a$.
 
Our results have some implications for the effects of lattice or molecular vibrations on the quantum electronic transport through atomic wires or
molecular junctions coupled to metallic leads.
Indeed, if one assumes that the electrical conductance of a quantum conductor is determined by its scattering properties~\cite{nazarov09}, 
such as the Landauer formula $G=(e^2/h) T$ in the simplest case, the transmission coefficients provide some information on the transport
properties. In the weak-coupling adiabatic limit, we have found that $T \sim 1/L_H$ and thus the resistance $1/G$ increases linearly
with the wire length $L_H$, as observed experimentally in macroscopic resistors.
This is also consistent with the linear increase of the dissipated energy with the wire length in that limit.
In the anti-adiabatic regime, there is no dissipation and we have found transmission resonances that do not depend
significantly on $L_H$. In particular, at the resonances with $T=1$ the EPC wire is an ideal conductor .
However, we also find transmission blockades $T=0$ corresponding to a perfect insulator.
More generally, the dissipated energy and the transmission probability vary significantly and independently with the parameters
$\omega_0, \gamma$, and $L_H$. Our numerical data and the exact result~(\ref{eq:reflection_dissipation}) suggest that both quantities are related
but we do not understand the precise relation yet.
In particular, the strict classical relation between (macroscopic) dissipation and resistance, $E_D \propto G^{-1}$ does not seem to be always fulfilled
in the present microscopic model.

Another open question is the effect of the effective interaction between electrons that is induced by the electron-phonon coupling, if more than one electron
is in the EPC structure at the same time.
The TEBD method can be used to determine the linear and nonlinear transport properties of interacting wires coupled to leads with a finite density
of charge carriers~\cite{einh12}.  Thus we plan to extend the present study of EPC wires to the case of finite electronic density.

\begin{acknowledgments}
We thank Florian Dorfner, Fabian Heidrich-Meisner, Kyle Poland, and Andreas Weichselbaum for helpful discussions.
We acknowledge support from the DFG (Deutsche Forschungsgemeinschaft) through
grant Nos.~JE~261/2-1 in the Research Unit
\textit{Advanced Computational Methods for Strongly Correlated Quantum Systems} (FOR 1807).
Some calculations were carried out on the cluster system at the Leibniz Universit\"{a}t Hannover.
\end{acknowledgments}

\bibliographystyle{biblev1}
\bibliography{references3}

\end{document}